\documentstyle[aps,prl,multicol]{revtex}
\input{psfig}

\begin{document}
\renewcommand{\Re}{\, {\rm Re}}
\renewcommand{\Im}{\, {\rm Im}}
\newcommand{\CC}{{\rm C.C.}}
\newcommand{\vecq}{{\bf q}}
\newcommand{\vecR}{{\bf R}}
\newcommand{\vecF}{{\bf F}}
\newcommand{\vecA}{{\bf A}}
\newcommand{\vecB}{{\bf B}}
\newcommand{\vecr}{{\bf r}}
\newcommand{\vecy}{{\bf y}}
\newcommand{\veck}{{\bf k}}
\newcommand{\vecv}{{\bf v}}
\newcommand{\vecE}{{\bf E}}
\newcommand{\vecj}{{\bf j}}
\newcommand{\kperp}{{{\bf k}_\perp}}
\newcommand{\sinch}{{\rm sinch}}
\newcommand{\cth}{{\rm cth}}
\newcommand{\bartau}{{{\bar \tau}_T}}
\renewcommand{\Re}{{\rm Re}}
\newcommand{\spc}{{\,\,\,\,\,\,\,\,}}
\newcommand{\bea}{\begin{eqnarray}}
\newcommand{\eea}{\end{eqnarray}}
\renewcommand{\[}{\begin{equation}} 
\renewcommand{\]}{\end{equation}}
\newcommand{\bef}{\begin{figure}} 
\newcommand{\ef}{\end{figure}}
\newcommand{\ie}{{\it i.e.}}
\newcommand{\eg}{{\it e.g.}}
\newcommand{\llabel}[1]{\label{#1}}
\newcommand{\eq}[1]{Eq.~(\ref{#1})} 
\newcommand{\fig}[1]{Fig.~\ref{#1}} 

\def\Journal#1#2#3#4{{#1} {\bf #2}, #3 (#4)}

\def\NCA{Nuovo Cimento}
\def\NIM{Nucl.\ Instrum.\ Methods}
\def\NIMA{{Nucl.\ Instrum.\ Methods}A}
\def\NPB{{Nucl.\ Phys.\ } B}
\def\PLB{{Phys.\ Lett.\ } B}
\def\PRL{Phys.\ Rev.\ Lett.\ }
\def\PRD{{Phys.\ Rev.\ } D}
\def\PRB{{Phys.\ Rev.\ } B}
\def\APL{{Appl.\ Phys.\ Lett.\ }}
\def\ZPC{{Z.\ Phys.\ } C}
\def\SSC{{Solid State Commun.\ } C}
\def\ANP{{Ann.\ Phys.\ } C}


\title{Magnetotransport of coupled electron-holes}
\author{Y. Naveh$^1$ and B. Laikhtman$^2$}
\address{$^1$Department of Physics and Astronomy, 
State University of New York, Stony Brook, NY 11794-3800, USA\\
$^2$Racah Institute of Physics, Hebrew University, Jerusalem 91904, Israel} 
\maketitle


\begin{abstract} 
The carriers in InAs-GaSb double quantum wells are hybrid
``electron-holes''. We study the magnetotransport properties of such
particles using a two-component Keldysh technique, which results in a
semi-analytic expression for the small-field current. We show that zero
temperature current can be large even when the Fermi energy lies
within the hybridization gap, a result which cannot be understood
within a semiclassical (Boltzmann) approach. Magnetic field dependence
of the conductance is
also affected significantly by the hybridization of electrons and
holes.
\end{abstract}

\begin{multicols}{2}

The concept of electrons and holes as the basic quasiparticles in
semiconductor physics is usually very useful for the study of
transport properties. However, in InAs-GaSb heterostructures this
concept breaks down. The reason for this is that electrons in the
InAs layer can be isoenergetic with holes in the 
GaSb layer. Then, hybridization occurs between conduction and valence
band states\cite{Altarelli 83}, and the single-particle 
excitations in the system are
hybrid electron-holes\cite{Naveh 95}. The effect of hybridization on
in-plane transport in the case where the carriers are of the same type
was studied in the past\cite{Palevski 
90,Vasko 93,Berk 94+95}, and was shown to result in ``resistance
resonance'' if the scattering times of carriers in the two wells are
not the same. Hybridization of electrons in two wells also
significantly changes their plasmon spectrum\cite{Gumbs 95,Das Sarma 98}.
However, the case of electron-hole hybridization is very
different. For example, in contrast to the electron-electron case, it is not
immediately clear what charge is carried by the electron-hole excitations. 
In addition, those excitations follow a strongly
non-monotonic dispersion law\cite{Naveh 95,Quinn 96,de-Leon 99}, and a
relatively large hybridization gap (later referred to as plain 'gap')
is formed. As a result, all transport properties are affected
by hybridization even if the scattering rates in the wells are the
same.  

Much experimental effort has been put into studying these particles.
The existence of the gap was confirmed\cite{Yang 97,Lakrimi 97} and
subsequent measurements have revealed more details of the
non-monotonic spectrum\cite{Marlow 99,Poulter 99,Vasilev 99}. Recent
transport measurements\cite{Cooper 98,Yang 99} have shown strong
resistance resonance. However, it was shown\cite{Cooper 98,Yang 99}
that the conductance remains finite even 
when the Fermi level is in the gap. In this regime current carriers are
the hybrid electron-holes.  

Analyzing theoretically the transport
properties of these hybrid particles requires a much more basic approach than
in the calculation of the spectrum. This is because natural broadening
of levels around the gap plays a crucial role in determining the
transport at relevant energies, and, as will be shown below, is the reason for
the finite in-gap current.  Previous theoretical works studying the
transport of particles obeying non-monotonic dispersion
laws\cite{Shvartsman 83} where based on semiclassical assumptions, and
thus could not produce this result.  Below we present a full quantum
theory for the magnetotransport of hybrid electron-holes. For the sake
of definiteness we consider an InAs/GaSb double quantum well system.

Our theory is based on a two component Keldysh technique.
In the absence of scattering and at zero electric and magnetic fields,
the Hamiltonian for the system takes the form\cite{Laikhtman 97} 
\[ \llabel{Hamiltonian}
H_0 = \left( 
\begin{array}{cc}
E_{e k} & -(i k_x + k_y)w \\
(i k_x - k_y) w & E_{h k} 
\end{array}
\right),
\]
where $E_{ek} = \hbar^2 k^2/2m_e$, $E_{hk} = E_{g0} - \hbar^2 k^2/2m_h$,
$\veck$ is the in-plane wavevector, $m_{e,h}$ are the in-plane electron and
hole effective masses, and $E_{g0} > 0$ is the energy difference between the
first hole and electron levels. Under applied electric and magnetic
fields, the Hamiltonian takes the same form
as in (\ref{Hamiltonian}), but with the substitution
\[
E_{\alpha k} \rightarrow E_{\alpha k} - e \vecF \cdot \vecr,
\,\,\,\,\,\,\,\, \veck \rightarrow -i\nabla - {e \over \hbar c} \vecA(\vecr),
\]
with $\vecF$ the electric field, $\vecA$ the vector potential, and
$\vecr$ the in-plane position. 

In what follows we calculate the current by first obtaining an
explicit expression for the density matrix
\[	\llabel{rhoG}
\rho({\bf r}_{1}, {\bf r}_{2}) =
        \int G^{-+}({\bf r}_{1}, {\bf r}_{2}, E) \, {dE \over 2\pi i},
\]
with $G$ the Keldysh Green's function\cite{Lifshits
81}. Note that $G$ here has 16 
components, 4 Keldysh components for each of the 4 terms
in the Hamiltonian. It satisfies the Dyson
equation\cite{Lifshits 81}
\begin{mathletters}		\llabel{Dyson}
\begin{eqnarray}
{\hat G}_{01}^{-1} G_{12} = \sigma_z + \sigma_z \Sigma_{13} G_{32},
\label{eq:evol.6a} \\
G_{12} {\hat G}_{02}^{-1 \ast} =
              \sigma_z + G_{13} \Sigma_{32} \sigma_z,
\end{eqnarray}
\end{mathletters}
\noindent
where $\Sigma$ is the self energy, $\sigma_z$ the Pauli matrix with
respect to Keldysh indices, and 
\[
{\hat G}_{0j}^{-1} = i\hbar {\partial\over \partial t_{j}} -
        H_{jj},
\]
in which $\partial / \partial t_j$ operates on the $j$th variable of
$G_{12}$.

In the presence of magnetic field $G$, $\Sigma$, and $\rho$ 
are not translationally invariant. We therefore work with the
translationally invariant functions\cite{Laikhtman 94}
\bea 	\llabel{translation}
\tilde{G}& & ({\bf r}_{1} - {\bf r}_{2}, E) \nonumber \\
 & & =
        G({\bf r}_{1}, {\bf r}_{2}, E - e\vecF \cdot \vecR)
        \exp    \left[
         -{i e \over \hbar c} \phi ({\bf r}_{1}, {\bf r}_{2})
                \right],
\eea
and similar definitions for $\tilde \Sigma$ and $\tilde \rho$. Here
$\vecR = (\vecr_1 + \vecr_2) / 2$, 
\[
\phi ({\bf r}_{1}, {\bf r}_{2}) =
        - {1 \over 2} {\bf B} \cdot ({\bf r}_{1} \times {\bf r}_{2}) +
        f({\bf r}_{1}) - f({\bf r}_{2}),
\label{eq:evol.ts.9}
\]
${\bf B}$ is the magnetic field, and $f(\vecr)$ is the gauge function,
defined by $\vecA(\vecr) = {1 \over 2}\vecB \times \vecr + \nabla f(\vecr)$
\cite{Laikhtman 94}. Furthermore, in electric field even $\tilde
G$ is not translationally 
invariant. This complication can be neglected if the energy
of the carriers and their wavefunctions are not affected appreciably by
the field. The restriction on the field is thus
\[	\llabel{smallF}
{e F \tau_{e,h} \over \hbar k_F}, \,\,\,  {e F \over k_F \Delta } \ll 1,
\]
with $\tau_{e,h}$ the elastic scattering times of electrons and holes, 
$k_F$ the Fermi wavenumber, and
$\Delta = w k_0$ the hybridization gap ($k_0$ is defined by $E_{ek_0}
= E_{hk_0}$).
When interested in magnetic phenomena this also requires
\[	\llabel{smallFB}
{e F R_L \tau_{e,h}\over \hbar}, \,\,\,  {E_{c e,h} \over E_F} \ll 1,
\]
with $R_L$ the Larmor radius and $E_{c \alpha} = \hbar e B / m_\alpha c$. In
practice, these
limitation are not strong and are applicable to a wide range of experiments. 

We now expand Eqs.~(\ref{Dyson}) for all their components, use
\eq{translation}, and 
transform into $\veck$-space. The resulting equations
for the $(-+)$ components of $\tilde G$ take forms similar to the
evolution
equation (4.6) of Ref.~\cite{Laikhtman 94}, but which combine the
various $e$-$h$ components of $\tilde G$, and with collision integrals
given by
\end{multicols}
\begin{mathletters} 	\llabel{collint}
\begin{eqnarray}
I_{\alpha\alpha{\bf k}} & = &
                \left(
        \tilde{\Sigma}_{\alpha\alpha{\bf k}}^{r} - \tilde{\Sigma}_{\alpha\alpha{\bf k}}^{a}
                \right)
        \tilde{G}_{\alpha\alpha{\bf k}}^{-+} -
        \tilde{\Sigma}_{\alpha\alpha{\bf k}}^{-+}
                \left(
        \tilde{G}_{\alpha\alpha{\bf k}}^{a} - \tilde{G}_{\alpha\alpha{\bf k}}^{r}
                \right) ,  
\label{eq:evol.cl.5a} \\
I_{{\bf k}\alpha\beta} & = &
        \tilde{\Sigma}_{\alpha\alpha{\bf k}}^{r} \tilde{G}_{\alpha\beta{\bf k}}^{-+} -
        \tilde{\Sigma}_{\alpha\alpha{\bf k}}^{-+} \tilde{G}_{\alpha\beta{\bf k}}^{a} +
        \tilde{G}_{\alpha\beta{\bf k}}^{r} \tilde{\Sigma}_{\beta\beta{\bf k}}^{-+} -
        \tilde{G}_{\alpha\beta{\bf k}}^{-+} \tilde{\Sigma}_{\beta\beta{\bf
        k}}^{a},
\label{eq:evol.cl.5c} 
\end{eqnarray}
\end{mathletters}
\vspace{-0.6cm}
\begin{multicols}{2}
\noindent
where $\tilde G^r = \tilde G^{--} - \tilde G^{-+}$, $\tilde G^a =
\tilde G^{--} - \tilde G^{+-}$, $\tilde \Sigma^r = \tilde \Sigma^{--}
+ \tilde \Sigma^{-+}$, and $\tilde \Sigma^a = \tilde \Sigma^{--}
+ \tilde \Sigma^{+-}$, and where we omitted the explicit dependence on
energy for brevity.
We also use the convention that if both $\alpha$ and $\beta$
appear in the same equation, they are distinctly different (\ie, if
$\alpha = h$ then $\beta = e$ and vice versa).

We now make the physical assumption that scattering in
the two wells is not correlated. Formally, this means that $\tilde
\Sigma$ is 
diagonal. Also, we follow the usual practice and neglect the small
renormalization of the 
spectrum due to scattering. In this case $\tilde\Sigma$ becomes purely
imaginary and 
can be written in the form
\[ \label{SigmaGamma} 
\tilde{\Sigma}_{\bf k}^{a}(E) = - \tilde{\Sigma}_{\bf k}^{r}(E) =
                i \left(
\begin{array}{cc}
        \Gamma_{ek}(E)      & 0     \\
\displaystyle
        0       & \Gamma_{hk}(E)
\end{array}     \right).
\]
$\Gamma_{\alpha k}(E)$ are found within the self-consistent Born
approximation by writing down the general 
expression for the self energy,
\[ 	\llabel{SigmaGu}
\tilde\Sigma_{\alpha\alpha \veck}(E) = \int\tilde G_{\alpha\alpha
\veck-\vecq}(E) u_{
\alpha {\bf q}} \,
        {d^{2}{\bf q}\over(2\pi)^{2}}, 
\]
where $u_{\alpha {\bf q}}$ are the elastic scattering matrix elements
squared. $\tilde G_{\alpha\alpha \veck}$ is then found by solving the general
relation\cite{Laikhtman 94}
\[ \label{GSigma}
E  \tilde G^{r,a} - {1 \over 2}  \{ H, \tilde G^{r,a}\}
        = 1 + {1 \over 2}  \{ \tilde \Sigma^{r,a}, \tilde G^{r,a} \},    
\]
with $\{ A,B \} = AB + BA$. The solution of this equation is
\end{multicols}
\[ 	\llabel{GGamma}
\tilde{G}_{k}^{r}(E) = \tilde{G}_{k}^{a\dagger}(E) = {1 \over R(E, k)}
                \left(
\begin{array}{cc}
        E - E_{hk} + i\Gamma_{hk}       & - w(ik_{x} + k_{y})   \\
        w(ik_{x} - k_{y})               & E - E_{ek} + i\Gamma_{ek}
\end{array}     \right),
\label{eq:col.9a} 
\]
with 
\[
R(E, k) = (E - E_{ek} + i\Gamma_{ek})(E - E_{hk} + i\Gamma_{hk}) -
        w^{2}k^{2},
\]
\begin{multicols}{2}
\noindent
and where we used Eqs.~(\ref{Hamiltonian}) and (\ref{SigmaGamma}). 
In the same approximation, substituting \eq{GGamma} in
\eq{SigmaGu} gives
\[	\llabel{SigmaKu}
\tilde\Sigma_{\alpha\alpha \veck}^{r}(E) = 
        -i \int \Im K_{1 1}^{\beta}(E, q)  u_{\alpha {\bf k-q}} \,
        {d^{2}{\bf q}\over(2\pi)^{2}},
\label{eq:col.13a} 
\]
where the family of kernels $K_{pq}^\alpha$ are defined by
\[ \llabel{kernels}
K^\alpha_{pq}(E, k) = {\left[ E - E_{\alpha k} - i\Gamma_{\alpha k}(E)
\right]^p \over
\left[ R(E, k)^{q}\right]^\ast}.
\]
For short-range scattering $u_{\alpha \veck}$, and therefore $\tilde
\Sigma$ and $\Gamma$, do not depend on
$\veck$. Its value can be found from
Eqs.~(\ref{SigmaKu},\ref{kernels}) at energies far from 
the gap, where they are expressed in terms of the relaxation times in
non-coupled wells,
\[	\llabel{tauu}
{\hbar \over \tau_\alpha} = {u_\alpha \over 2} \int_0^\infty \delta(E -
E_{\alpha k}) k \, dk = {m_\alpha u_\alpha \over 2 \hbar^2}.
\]
Combining Eqs.~(\ref{SigmaGamma},\ref{SigmaKu},\ref{tauu}) results in
two self consistent equations for $\Gamma_e$, $\Gamma_h$,
\[	\llabel{Gammaselfcons}
\Gamma_{\alpha}(E) = {\hbar^3 \over  \pi m_\alpha \tau_\alpha}
        \int_0^\infty \Im  K_{1 1}^{\beta}(E, k)   k \, dk.
\]

Next, we substitute Eqs.~(\ref{SigmaGamma}) and (\ref{GGamma}) into
\eq{collint} which results in collision integrals depending only on
the $(-+)$ components of $\tilde G$ and $\tilde \Sigma$. We then linearize
the equations for these components with respect to the electric field,
using the unperturbed (zero-order) 
function $\tilde G^{-+(0)} = f_0(E) \langle \tilde
G^a - \tilde G^r 
\rangle$\cite{Laikhtman 94}, with $f_0(E)$ 
the Fermi function.  After the linearization the
equations for the Green's functions are integrated with respect to 
energy. With the use of \eq{rhoG} this leads to four transport
equations for the components of the density 
matrix.  The solution to these equations is
found in a straightforward, though tedious, manner. It
is given in terms of the coefficients
$\vec{\chi} (k)$, defined by
\begin{mathletters}
\llabel{rhochi}
\begin{eqnarray}
\tilde \rho_{\alpha\alpha}^{(1)} (\veck) & = & \vec{\chi}^{\alpha\alpha}(k) 
\cdot \veck, \\
\tilde \rho_{\alpha\beta}^{(1)} (\veck) & = & \vec{\chi}^{\alpha\beta}(k) 
\cdot \veck,
\eea
\end{mathletters}
\noindent 
with $\tilde \rho^{(1)}$ the term of $\tilde \rho$ linear in $F$. The
solution reads
\end{multicols}
\begin{mathletters}	\llabel{chi}
\bea
\chi^{\alpha\alpha}_x(k)  & = & \pm {2 e\hbar^2 F}     {\omega_{\beta k}
        \zeta_{\alpha}(k) -
        \omega_{k}
        \zeta_{\beta}(k)  \over
\omega_{e k} \omega_{h k} - \omega^2_{k}}, \\
\chi^{\alpha\alpha}_y(k)  & = & \pm {1 \over 2}   {E_{c\beta}
\gamma_k \chi^{\beta\beta}_x(k) - E_{c\alpha}
\gamma_{\beta k} \chi^{\alpha\alpha}_x(k) \over \gamma_{e k} \gamma_{h
k} -
\gamma_k^2}, \\
2 \Re \chi^{eh}_x & = & {2 m_r w^2 \Gamma \over E_k^2 + \Gamma^2} \left\{ \xi_2^h
(k) - \xi_2^e (k) 
  + {1 \over \hbar^2} \left[ {E_{c h} \Gamma_{e} \over 
2 \bar\gamma_k^2} + {E_k \over \Gamma} \right] \chi^{hh}_{x} (k) - {1
\over \hbar^2} \left[ {E_{c e} 
\Gamma_h \over 
2 \bar\gamma_k^2} + {E_k \over \Gamma} \right] \chi^{ee}_x (k) \right\}, \\
2 \Re \chi^{eh}_{y}  & = & 0,
\label{eq:lin.sol.7}
\eea
\end{mathletters}
\noindent
with the upper sign for $\alpha = e$ and the lower for $\alpha = h$.
Here
\[
\zeta_\alpha(k) = \xi_1^\alpha(k) + w^2 k^2 \eta(k),
\]
\[	\llabel{eta}
\eta(k) = 
        {1\over \pi m_{r}} 
        \int f_{0}(E)
                \Im K_{02}(E, k) \, dE +
        {2E_{k} 
        \over E_k^{2} + \Gamma^{2}} \
        \int f_{0}(E)
                \left[
        {1\over \pi m_{e}}
                \Im K^h_{12}(E, k)
        - {1\over \pi m_{h}} \Im K^e_{12}(E, k) \right] \, dE,
\]
\begin{multicols}{2}
\[
\xi_1^{\alpha}(k) = {-1\over \pi m_{\alpha}}
        \int f_{0}^{\prime}(E)              
        \Im K^\beta_{1 1}(E, k)
        \, dE,
\label{eq:lin.rhs.17a} 
\]
\[	\llabel{xi2}
\xi_2^{\alpha}(k) =  - {e F \over \pi m_{\alpha}} 
\int f_{0} (E) \Im K_{1 2}^{\beta} (E,k) \, {dE,}
\llabel{xibare} \\
\]
\noindent
\bea
\omega_{\alpha k} & = & 
        4\gamma_{\alpha k} +
{E_{c\alpha}^{2}\gamma_{\beta k}
        \over \gamma_{e k} \gamma_{h k} - \gamma_k^2}, \\
\omega_{k} & = &
        4\gamma_k + {E_{c e}E_{c h}\gamma_k
        \over \gamma_{e k} \gamma_{h k} - \gamma_k^2},
\eea
$\gamma_k = w^2 k^2 \Gamma / (E_k^2 + \Gamma^2)$, $\gamma_{\alpha k} =
\Gamma_\alpha + \gamma_k$, $\Gamma = \Gamma_e + \Gamma_h$, $E_k = E_{h
k} - E_{e k}$, $1/m_r = 1/m_e + 1/ m_h$, and $x$ is in the direction
of the electric field. 

The current density is given by
\[
\vecj = {e i \over S \hbar} \langle
[H, \vecr] \rangle
\]
with $S$ the area of the sample. 
Performing the average over the density matrix $\tilde \rho$, and using
\eq{rhochi} results in
\[	\llabel{jchi}
\vecj =  {e \hbar k_F^3 \over 2 \pi} \int dk \, \left[
{{\vec\chi}^{ee}(k) \over m_e} + {{\vec\chi}^{hh}(k) \over m_h} + 2
\Re {{\vec\chi}^{eh}(k) \over m_r} \right].
\]
The mixing of electrons and holes is affecting both the diagonal
contributions 
to the current [see Eq.~\ref{chi}(a,b)], and the unusual off-diagonal
contribution (the origin of the latter is solely due to this mixing).

Equations (\ref{chi},\ref{jchi}) present a semi-analytic expression for the
current at any temperature and for any value of $\Gamma$, provided the
conditions (\ref{smallF},\ref{smallFB}) are met. [The only numerical
tasks required in order to calculate the current are the integration
over energy in Eqs.~(\ref{eta}--\ref{xi2}) and the solution of
Eqs.~\ref{Gammaselfcons}].
Results for the sheet conductance at zero magnetic field and zero
temperature as a function of Fermi energy are shown in
Fig.~1. Magnetoconductance curves for the case where the Fermi level
is in midgap are shown in Fig.~2. 

Two important conclusions can be drawn from the results shown in
Figs.~\ref{1conductance},\ref{2magnetoconductance}. First, it is clear
that the zero-temperature conductance does not vanish even when the
Fermi level is in the gap. This is due to the finite broadening of the
levels $\Gamma$. Moreover, as long as the particles are not localized,
conductivity is finite even for very small $\Gamma$.  This is because
the scattering mean free time (proportional to the conductivity)
scales as $1 / \Gamma$ while the overlap of levels in the two
hybridized bands is linear with $\Gamma$.  In the case when $\Gamma \ll \Delta \ll
E_{g0}$, Eqs.~(\ref{chi},\ref{jchi}) result in a crude estimate of the
conductivity, $g
\approx (e^{2}/h)(E_{g0} / \Delta)$. 
At very small values of $\Gamma/\Delta$ states in
the middle of the gap become 
localized, and the conductivity falls to a value close to $e^2/h$
(this effect cannot be described by the first Born approximation
used in this work). It
would be interesting to study the competition between this 
quantum resistance, usually proportional to $\ln(L/l)$, with $L$ the
sample length and $l$ the mean free path, and our hybridization resistance,
proportional to $\Delta / E_{g0}$. An opportunity to study the two
terms is unique to our system because $\Gamma$ does not play a role in
the conductivity due to states outside the mobility gap.  In addition, the
logarithmic dependence of the 
quantum correction on $L/l$ was obtained only for purely metallic
systems. In our case the Fermi level is located in the hybridization
gap, and this 
correction may assume a different form. In any case, it is emphasized
that our result of finite conductivity in the gap, which is consistent
with recent experimental results \cite{Cooper 98,Yang 99}, cannot be
understood within a semiclassical Boltzmann theory.

\vspace{-0.7cm}
\begin{figure}[h]
\narrowtext
\centerline{\hspace{-3.0cm}}
\hspace{-2cm}
\psfig{figure=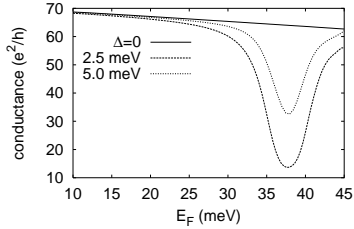,angle=-90,height=1.3in}
\caption{Conductance as a function of Fermi energy for various
coupling strengths at zero temperature and zero magnetic
field. $\Delta = w k_0$ is the hybridization gap.
Here
$\hbar/\tau_e = 0.8$ meV, $\hbar/\tau_h = 0.6$ meV, and $E_{g0} = 50$ meV.}
\llabel{1conductance}
\end{figure}

\vspace{-0.7cm}
\begin{figure}[h]
\centerline{\hspace{-5.0cm}}
\hspace{-2.0cm}
\psfig{figure=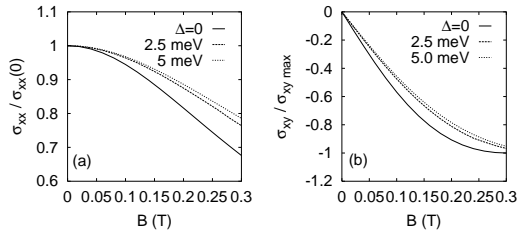,angle=-90,height=1.3in}
\caption{(a) longitudinal and (b) transverse conductance as a function
of magnetic field for the case when $E_F$ is at midgap and at zero
temperature. $\sigma_{xy \,\,{\rm max}}$  is the maximum value of
$\sigma_{xy}$ occurring at $B \approx 0.3 - 0.5$ T. $\tau_e$,
$\tau_h$, and $E_{g0}$ are as in Fig.~1.}
\llabel{2magnetoconductance}
\end{figure}

\vspace{-0.0cm}
Second, it is seen that magnetoconductance curves are affected by the
presence of electron-hole coupling. This result may be of 
practical importance in the characterization of InAs-GaSb samples. That is
because the usual characterization routines use sensitive fitting
procedures of classical 
magnetoconductance expressions to measured data in order to extract
four independent parameters: the electron and hole densities and
mobilities \cite{Mikhailova 94}. The corrections to the classical expressions
(i.e., to the case $\Delta = 0$) due to hybridization may significantly affect
the apparent value of these parameters.

In summary, we have studied the transport properties of coupled
electron-holes within a full quantum approach. We arrived at a
semi-analytic expression for the current under weak electric
and magnetic fields, Eqs.~(\ref{chi},\ref{jchi}). This current 
shows strong resistance resonance, does not vanish in the gap, 
and its
magnetic-field dependence depends on the strength of
coupling.

We are grateful to O. Agam, I. E. Aleiner, L. J. Cooper, N. K. Patel,
M. Finkelstein, and 
L. D. Shvartsman for valuable discussions. The work in Jerusalem was 
supported by the Israel Science Foundation founded by the
Israel Academy of Sciences and Humanities.

\end{multicols} 
\end{document}